\documentclass[prd,superscriptaddress,showpacs,preprintnumbers,amsmath,amssymb]{revtex4}

\usepackage{graphicx}

\def\be{\begin{equation}}
\def\ee{\end{equation}}
\def\bea{\begin{eqnarray}}
\def\no{\noindent}
\def\eea{\end{eqnarray}}
\def\beq{\begin{eqnarray}}
\def\eeq{\end{eqnarray}}

\begin{document}
\title{Delicate f(R) gravity models with disappearing cosmological constant
and observational constraints on the model parameters}

\author{Abha Dev}
\affiliation{Miranda House, University of Delhi, Delhi 110 007, India}
\author{Deepak Jain}
\affiliation{ Deen Dayal Upadhyaya College, University of Delhi, New
Delhi 110 015, India }
\author{S. Jhingan}
\affiliation{Centre for Theoretical Physics, Jamia Millia Islamia,
New Delhi, India}
\author{S. Nojiri}
 \affiliation{Department of Physics, Nagoya University, Nagoya 464-8602, Japan}
\author {M. Sami}
 \affiliation{Centre
for Theoretical Physics, Jamia Millia Islamia, New Delhi, India}
\author{I. Thongkool}
\affiliation{Centre for Theoretical Physics, Jamia Millia Islamia,
New Delhi, India}

\begin{abstract}
We study the $f(R)$ theory of gravity using metric approach. In
particular we investigate the recently proposed model by Hu-Sawicki,
Appleby $-$ Battye and Starobinsky. In this model, the cosmological
constant is zero in flat space time. The model passes both the Solar
system and the laboratory tests. But the model parameters need to be
fine tuned to avoid the finite time singularity recently pointed in
the literature. We check the concordance of this model with the
$H(z)$ and baryon acoustic oscillation data. We find that the model
resembles the $\Lambda$CDM at high redshift. However, for some
parameter values there are variations in the expansion history of
the universe at low redshift.
\end{abstract}


\maketitle

\section{Introduction}

\vskip 0.5cm \noindent It is remarkable that different data sets of
complementary nature such as supernovae, baryon oscillations, galaxy
clustering, microwave back ground and weak lensing all taken
together strongly support the late time acceleration of
universe. In the standard lore, one assumes that the history of
universe is described by the general relativity (GR). The late time
acceleration can easily be captured in this frame work by
introducing a scalar field with large negative pressure known as
{\it dark energy} \cite{review}. In view of the fine tuning problem,
the scalar field models, specially those with tracker like
solutions, are more attractive compared to the models based on
cosmological constant. At present, observations are not in a
position to reject or to establish the dark energy metamorphosis. A
host of scalar field models have been investigated in the
literature. The scalar field models can fit the data but lack the
predictive power. It then becomes important to seek the support of
these models from a fundamental theory of high energy physics.

\vskip 0.5cm \noindent One can question the standard lore on
fundamental grounds. We know that gravity is modified at small
distance scales; it is quite possible that it is modified at large
scales too where it has never been confronted with observations
directly. It is therefore perfectly legitimate to investigate the
possibility of late time acceleration due to modification of
Einstein-Hilbert action. It is tempting to study the string
curvature corrections to Einstein gravity amongst which the
Gauss-Bonnet correction enjoys special status. A large number of
papers are devoted to the cosmological implications of string
curvature corrected
gravity\cite{fr,NOS,KM06,TS,CTS,Neupane,Cal,Sami06,Annalen,Sanyal}.
These models suffer from several problems. Most of these models do
not include tracker like solution and those which do are heavily
constrained by the thermal history of universe. For instance, the
Gauss-Bonnet theory with dynamical dilaton might cause transition
from matter scaling regime to late time acceleration allowing to
alleviate the fine tuning and coincidence problems. However, it is
difficult to reconcile this model with
nucleosynthesis\cite{TS,KM06}. Another possibility of large scale
modification is provided by non-locally corrected gravity which
typically involves inverse of d'Alembertian of Ricci scalar. The
non-local construct might mimic dark energy; the model poses
technical difficulties and there has been a little progress in this
direction\cite{Wood}. The large scale modification may also arise in
extra dimensional theories like DGP model which contains self
accelerating brane. Apart from the theoretical problems, this model
is heavily constrained by observation.

\vskip 0.5cm \noindent On purely phenomenological grounds, one could
seek a modification of Einstein gravity by replacing the Ricci
scalar by $f(R)$. The $f(R)$ gravity models have been extensively
investigated in past five
years\cite{Nojiri:2003ft,review1,FRB,FRB1}. The $f(R)$ gravity
theories  giving rise to cosmological constant in low curvature
regime are plagued with instabilities and on observational grounds
they are not distinguished from cosmological constant. The recently
introduced models of $f(R)$ gravity by Hu-Sawicki and Starobinsky
(referred as HSS models hereafter) with disappearing cosmological
constant\cite{HS,star} have given rise to new hopes for a viable
cosmological model within the framework of modified gravity ($f(R)$
gravity model with similar properties is proposed in
Ref.\cite{ABTT}). These models contain Minkowski space time as a
solution in the low curvature regime which is an unstable solution.
In high curvature regime these models reduce to cosmological
constant. Both the first and the second derivatives of $f(R)$ with
respect to R are positive. The positivity of the first derivative
ensures that the scalar degree of freedom, a characteristic of any
f(R) theory, is not tachyonic where as the positivity of second
derivative tells us that graviton is not ghost thereby guaranteing
the stability. In Starobinsky parametrization\cite{star}, $f(R)$ is
given by, $f(R) = R + \Lambda \left \lbrack \left ( 1 + R^2/R_0^2
\right )^{-n} - 1 \right]$. The HSS models can evade solar physics
constraints provided that the model parameters are chosen properly.
An important observation has recently been made by by Appleby $-$
Battye and Forolov\cite{AppBatt,frolov} (see also
\cite{Nojiri:2008fk}). The minimum of scalaron potential which
corresponds to dark energy can be very near to $\phi=0$ or
equivalently $R=\infty$. As pointed out in Ref.\cite{Tsujikawa}, the
minimum should be near the origin for solar constraints to be
evaded. Hence, it becomes most likely that we hit the singularity if
the parameters are not fine tuned.

\vskip 0.5cm
\noindent In order to check whether the $f(R)$ gravity theory is cosmological
viable or not, it is necessary that this theory  must be compatible with
the observations. In this work, we study the Starobinsky model using the
data from the recent observations which include $H(z)$ , Hubble
parameter  at various red-shifts and the Baryon Acoustic Oscillation
(BAO) peak from Sloan Digital Sky Survey (SDSS).

\vskip 0.5cm \noindent This paper is organised as follows. In
Section II, we describe the general properties of the model
highlighting the fine tuning problem. The Friedmann equation and the
special cases of trace equation is studied in Section III. The
cosmological constraints from the recent observations are described
in Section IV. Finally, Section V contains the results and
discussions.

\section{Naturalness of the model}

\vskip 0.5cm \noindent In this section we shall revisit f(R) models
with disappearing cosmological constant. These models have potential
capability of being distinguished from $\Lambda$CDM and could lead
to a viable cosmological model. However, even at the background
level this class of models are fine tuned. In what follows we shall
explicitly bring out these features specializing to Starobinsky
parametrization.

\vskip 0.5cm \noindent The action of  $f(R)$ gravity is given
by\cite{review1},
\begin{equation}\label{action}
S = \int\left[\frac{f(R)}{16\pi G} + \mathcal{L}_m \right] \sqrt{-g}
\quad d^4 x,
\end{equation}
\noindent which leads to the following equation of motion
\begin{equation}\label{eq:freqn}
 f'R_{\mu\nu}-\nabla_{\mu\nu}{f'}+\left(\Box f' - \frac{1}{2}f\right)g_{\mu\nu}
=8\pi G T_{\mu\nu}.
\end{equation}
\noindent Here prime $(')$ denotes the derivatives with respect to
$R$. The f(R) gravity theories apart from a spin-2 object
necessarily contain a scalar degree of freedom. Taking trace of
Eq.(\ref{eq:freqn}) gives the evolution equation for the scalar
degree of freedom,
\begin{equation}\label{eq:frtrace}
 \Box f' = \frac{1}{3} \left ( 2f - f' R \right ) + \frac{8\pi G}{3} T.
\end{equation}
\noindent It would be convenient to define scalar function $\phi$ as
\begin{equation}
 \phi \equiv f' - 1,
\end{equation}
\noindent which is expressed through Ricci scalar once f(R)  is specified.

\vskip 0.5cm
\noindent We can write the trace equation (equation $(\ref{eq:frtrace})$)
 in the term of $V$ and $T$ as
\begin{equation}
 \Box \phi = \frac{dV}{d\phi} + \frac{8\pi G}{3} T.
\end{equation}
\vskip 0.5cm
\noindent The potential can be evaluated using the following relation
\begin{equation}\label{Effpot}
 \frac{dV}{dR} = \frac{dV}{d\phi}\frac{d\phi}{dR}= \frac{1}{3}
\left ( 2 f - f' R \right )
 f''.
\end{equation}

\vskip 0.5cm \noindent Recently Hu, Sawicki and Starobinsky proposed
a functional form of $f(R)$ with the desirable properties of a
viable model \cite{HS,star}. In this paper we shall consider the
model in Starobinsky parametrization:

\begin{equation}  \label{func}
f(R) = R + \lambda R_0 \left[\left(1+\frac{R^2}{R_0^2}\right)^{-n}
-1\right].
\end{equation}
\noindent Here n and  $\lambda$ are greater than zero. And $R_0$ is
of the order of presently observed cosmological constant, $\Lambda =
8\pi G \rho_{vac}$. The properties of this model  can be summarized
as follows:

\begin{enumerate}
\item \no In the absence of matter $R^{\mu}_{\nu}$ is always a solution, $%
\lim_{R\rightarrow 0} f(R) = 0$. However, as $f^{\prime \prime }<0$, flat
space time is unstable.

\item \no For $|R| \gg R_0$, $f(R) = R - 2 \Lambda (\infty)$. The
high-curvature value of the effective cosmological constant is
$\Lambda(\infty) = \lambda R_0/2$.

\item \no The stability conditions for the adopted model are
\begin{equation}  \label{stabi}
f^{\prime }(R) >0, \quad f^{\prime \prime }(R) >0.
\end{equation}
\end{enumerate}

\vskip 0.5cm
\noindent In the Starobinsky model the scalar field $\phi$ is given by
\begin{equation}\label{eq:phi}
 \phi(R) = -\frac{2n \lambda R}{R_0(1+\frac{R^2}{R_0^2})^{n+1}}.
\end{equation}

\vskip 0.5cm
\noindent We can compute $V(R)$ for a given value of $n$.
In case of $n = 1$, we have
\begin{eqnarray}
\frac{V}{R_0} &=& \frac{1}{24 \left ( 1 + y^2 \right )^4 } \left
\lbrace \left ( -8 - 40y^2 - 56y^4 -24y^6 \right ) \lambda + \left (
3y + 11y^3 + 21y^5 - 3y^7 \right ) \lambda^2
\right \rbrace \nonumber \\
&&- \frac{\lambda^2}{8}\tan^{-1}{y},
\end{eqnarray}
\noindent where $y=R/R_0$. In case and of $n = 2$ we have,
\begin{eqnarray}
\frac{V}{R_0} &=& \frac{1}{480 \left ( 1 + y^2 \right )^6 } \left
\lbrace \left ( -160 - 1280 y^2 - 2880y^4 - 2560y^6 - 800y^8 \right
) \lambda
\right . \nonumber \\
&& + \left . \left ( 105y + 595y^3 + 2154y^5 + 106y^7 + 595y^9 +
105y^{11} \right ) \lambda^2
\right \rbrace \nonumber \\
&& - \frac{35\lambda^2}{160}\tan^{-1}{y}.
\end{eqnarray}
\vskip 0.5cm
\noindent In the FRW background, the trace equation
(equation $(\ref{eq:frtrace})$) can
be rewritten in the convenient form
\begin{equation}\label{eq:scalarfield}
 \ddot{\phi} + 3H\dot{\phi} + \frac{dV}{d\phi} =\frac{8\pi G}{3} \rho.
\end{equation}

\begin{figure}[htp]
\centering
\includegraphics[width=100mm]{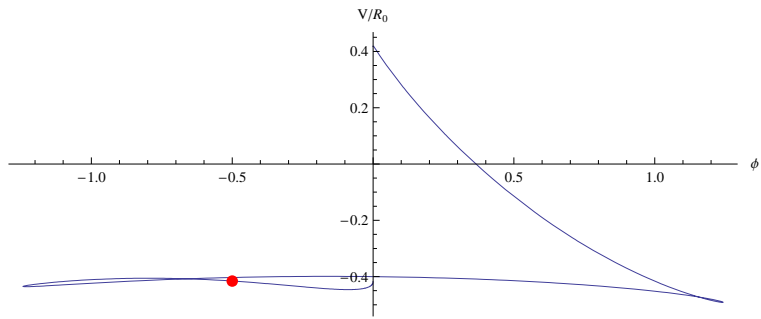}
\caption{Plot of the scalaron potential $V$ versus $\phi$ for $n =
2$ and $\lambda = 1.2$. The red spot marks the initial condition for
evolution of $\phi$.}\label{fig:n2l1p2v}
\end{figure}

\vskip 0.5cm
\no The time-time component of the equation of motion $(\ref{eq:freqn})$
gives the Hubble equation
\begin{equation}\label{eq:Friedmann}
 H^2 + \frac{d (\ln f')}{dt} H + \frac{1}{6}\frac{f - f'R}{f'}
= \frac{8\pi G}{3f'} \rho.
\end{equation}

\no We recover Einstein gravity in the limit $f^{\prime }=1$ . The
simple picture of dynamics which appears here is the following:
above infrared modification scale $(R_{0})$, the expansion rate is
set by the matter density and once the local curvature falls below
$R_{0}$ the expansion rate gets effect of gravity modification.

\vskip 0.5cm
\no For pressure less dust, the effective potential has
an extremum at
\begin{equation}  \label{extm}
2 f - R f^{\prime }= 8\pi G \rho.
\end{equation}
For a viable late time cosmology, the field should be evolving near
the minimum of the effective potential. The finite time singularity
inherent in the class of models under consideration severely
constraints dynamics of the field.
\subsection*{The finite time singularity and fine tuning of parameters}

\begin{figure}[htp]
\centering
\includegraphics[width=100mm]{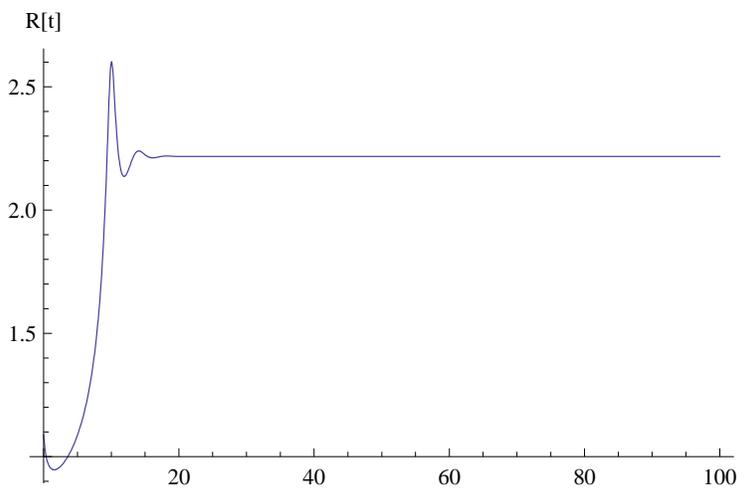}
\caption{The evolution of $R(t)$ in the case $n = 2$ and $\lambda =
1.2$. The initial condition is as marked by the red spot in figure
\ref{fig:n2l1p2v}.}\label{fig:n2l1p2r}
\end{figure}

\no The effective potential has minimum which depends upon $n$ and
$\lambda$. For generic values of the parameters, the minimum of the
potential is close to $\phi =0$, corresponding to infinitely large
curvature. Thus, while the field is evolving towards minimum, it can
easily oscillate to a singular point. However, depending upon the
values of parameters, we can choose a finite range of initial
conditions for which scalar field $\phi$ can evolve to the minimum
of the potential without hitting the singularity. For instance, let
us begin with the initial conditions as shown in figure
$\ref{fig:n2l1p2v}$ given by $\phi_{int} = -0.5$ ($R = 1.089$). We
choose $H = 1$ then $\dot{H}$ becomes $-1.818$ for a given value of
$R$ (see eq. $(\ref{eq:Friedmann})$). The minimum of the potential
for $n=2$ and $\lambda=1.2$ corresponds to $ R_{min} = 2.218$ or $
\phi_{min} = -0.051$ which is indeed close to singularity. The
numerical results are shown in figures \ref{fig:n2l1p2v} and
\ref{fig:n2l1p2r}.

\subsection*{The problem of large $n$, $\lambda$}

\begin{figure}[htp]
\centering
\includegraphics[width=100mm]{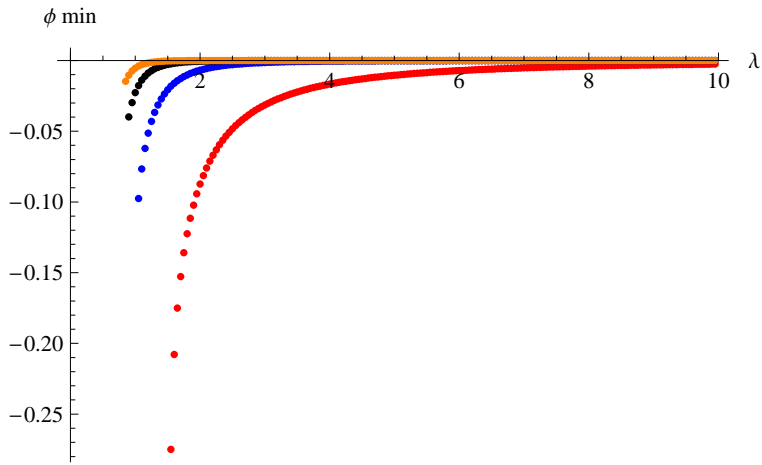}
\caption{Plot of $\phi_{min}$ versus $\lambda$ for different values
of $n$. With increase in $n$, $\phi_{min}$ moves towards zero
(corresponding to infinitely large value of $R$) for smaller values
of $\lambda$. The curves from bottom to top correspond to
$n=1,2,3,4$ respectively.}\label{fig:phmin}
\end{figure}

\vskip 0.5cm \no We find that the range of initial conditions
allowed for the evolution of $\phi$ to the minimum without hitting
singularity shrinks as the numerical values of parameters $n$ and
$\lambda$ increase. This is related to the fact that for larger
values of $n$ and $\lambda$, the minimum fast moves towards
$\phi=0$, see figure \ref{fig:phmin}.

\begin{figure}[htp]
\centering
\includegraphics[width=100mm]{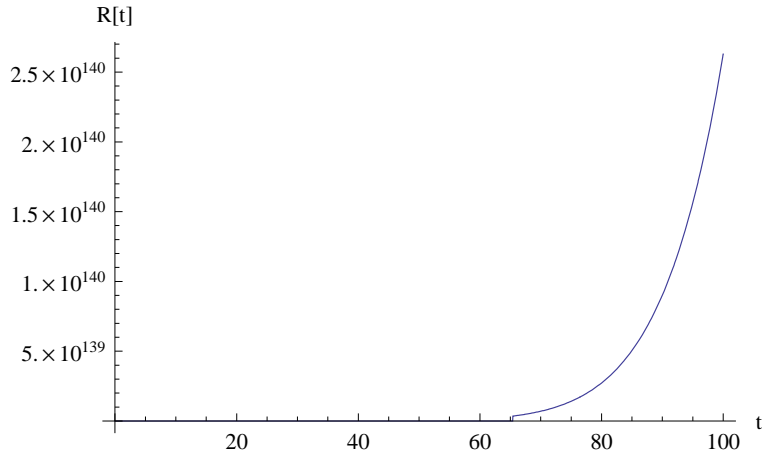}
\caption{The evolution of $R(t)$ for $n = 1$ and $\lambda =
4$.} \label{fig:n1l4r}
\end{figure}

\vskip 0.5cm \no For example, in the case of $n = 1, \lambda = 4$,
we find that $\phi_{min} = -0.016$ which is close to zero. The
initial value of $\phi$ which is quite near to minimum also gives
the divergence of $H(t)$ and $R(t)$ at the finite time (see Fig.
\ref{fig:n1l4r}).

\vskip 0.5cm \no On the other hand, one can see from the figure
\ref{fig:phmin} that $\phi_{min}$ also shifts to zero when $n$ is
increased. For $n=4$ and $\lambda =1.1$ we have $\phi_{min} =
-0.003$. The range of initial conditions for which the scalar field can
evolve to the minimum is very small.

\subsection*{Adding $\alpha R^2$ term and curing the singularity}
\vskip 0.5cm
\no We know that in case of large curvature, the quantum effects become
important leading to higher curvature corrections. Keeping this in
mind, let us consider the modification of Starobinsky's model,
\begin{equation}
f(R) = R + \frac{\alpha}{R_0} R^2 +R_0 \lambda \left \lbrack -1 +
\frac{1}{(1+\frac{R^2}{R_0^2})^n} \right \rbrack,
\end{equation}
then $\phi$ becomes
\begin{equation}\label{eq:phi2}
\phi(R) = \frac{R}{R_0}\left[2\alpha  -\frac{2n \lambda
}{(1+\frac{R^2}{R_0^2})^{n+1}}\right].
\end{equation}

\vskip 0.5cm
\no When $|R|$ is large the first term which comes from $\alpha R^2$
dominates. In this case, the curvature singularity $R = \pm \infty$
 corresponds to $\phi = \pm \infty$. Hence, in this modification, the
minimum of the effective potential is separated from the curvature
singularity by the infinite distance in the $\phi,V(\phi)$ plane.

\vskip 0.5cm \no For $n = 2$, $\phi$ and $V(\phi)$ are given by
\begin{eqnarray}
 \phi(y) &=& 2\alpha y - \frac{4 \lambda y }{(1 + y^2)^3}, \\
\frac{V}{R_0} &=& -\frac{1}{480(1+y^2)^6} \left \lbrace \lambda^2 y
\left ( -105 - 595y^2 -2154 y^4 + 106 y^6
+ 595 y^8 + 105 y^{10}\right ) \right \rbrace \nonumber \\
&& -\frac{1}{3(1+y^2)^3} \left \lbrace 1 + 5y^2 + \alpha \left ( 3 +
8y^2 + 9y^4 + 4y^6 \right) \right \rbrace
+ \frac{1}{3} \alpha y^2 \nonumber \\
&& + \frac{1}{32} \left (32 \alpha - 7\lambda \right ) \tan^{-1}(y).
\end{eqnarray}
\vskip 0.5cm \no For $n = 2, \lambda = 2$ and $\alpha = 0.5$, we
have a large range of the initial condition for which the scalar field
evolves to the minimum of the potential. Though the introduction of
$R^2$ term formally allows to avoid the singularity but can not
alleviate the fine tuning problem as the minimum should be brought
near to the origin to respect the solar constraints. Last but not
the least one could go beyond the approximation (see equation (\ref{extm}))
by iterating the trace equation and computing the corrections to $R$ given
by equation (\ref{extm}). As pointed by Starobinsky\cite{star}, such a
correction might become large in the past. This may spoil the thermal
history and thus needs to be fine tuned. The aforesaid
discussion makes it clear that HSS models are indeed fine tuned and
hence very delicate.

\begin{figure}[htp]
\centering
\includegraphics[width=100mm]{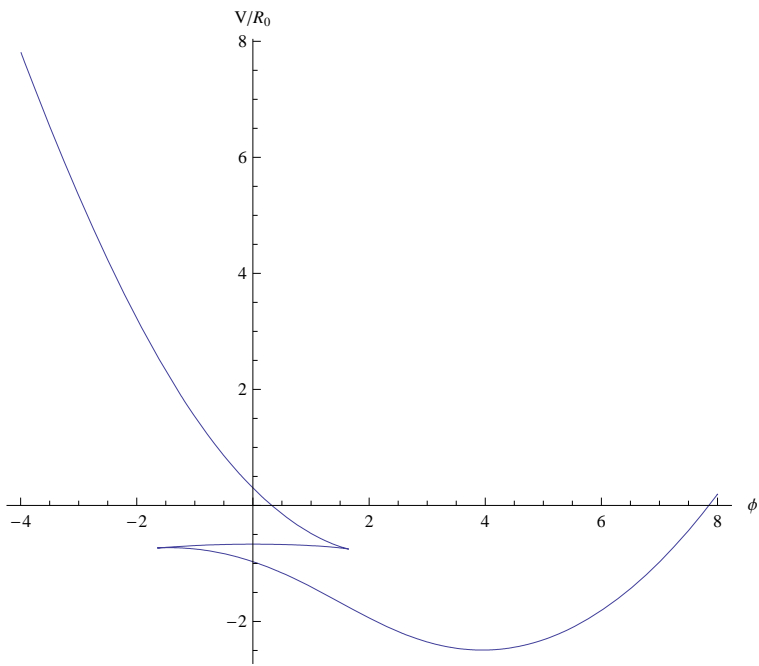}
\caption{Plot of the effective potential for $n=2$, $\lambda=2$ and
$\alpha=1/2$ in presence of $R^2$ correction. The minimum of the
effective potential in this case  is located at $\phi_{min}=3.952
\,\,(R_{min} = 3.958)$.}\label{fig:n2l2a0p5v}
\end{figure}

\section{Parametrization and Analysis}

\no The Friedmann equation can be casted in the form,
\begin{equation}\label{nfreid}
H^{2}=\frac{6\kappa \rho +Rf^{\prime }-f}{6(f^{\prime}+R_{,N}f^{''
})},
\end{equation}
where $N=\ln a$. To parameterise the system in a more convenient
form we define:
\begin{equation}\label{repar}
\kappa \rho =\Omega _{m}H_{0}^{2},\quad R_{0}=3\Omega _{\Lambda
}H_{0}^{2}, \quad x=\frac{R_{0}}{R}.
\end{equation}
Since $|R|\gg R_{0}$ we have  $x\ll 1$. The $f, f'$ and $f''$ for the
Starobinsky model are of the form
\begin{eqnarray}\label{fprime}
  \frac{f}{H_0^2} & = & 3 \Omega_{\Lambda} \left[\frac{1}{x}-
  \lambda +\lambda\left(1+\frac{1}{x^2}\right)^{-n} \right], \nonumber \\
  f' &=& 1-\frac{2n\lambda}{x} \left(1+\frac{1}{x^2}\right)^{-(1+n)}, \\
  H_0^2 f'' &=& \frac{2n\lambda}{3\Omega_{\Lambda}}
   \left[-1+\frac{(2n+1)}{x^2}\right]
   \left(1+\frac{1}{x^2}\right)^{-(2+n)}. \nonumber
\end{eqnarray}
\vskip 0.5cm
\no The trace equation can  be re-written as
\begin{equation}\label{treqn}
    \frac{1}{x} + 2 \lambda \left[ \left(1+\frac{1}{x^2}\right)^{-n} +
     \frac{n}{x^2}\left(1+\frac{1}{x^2}\right)^{-n-1} -1\right] =
    \frac{\Omega_{m}}{\Omega_{\Lambda}}.
\end{equation}
\no Here $\Omega_{m} = \Omega_{m0}(1+z)^3$ is the matter
density of the universe at a given redshift. $\Omega_{m0}$, is
present fractional matter density and $\Omega_{\Lambda}$, is the
present fractional density of the vacuum.

\vskip 0.5cm \no The Friedmann equation (equation (\ref{nfreid}))
can now be written as
\begin{equation}\label{friedman}
\frac{H^2}{H_0^2} =  \frac{f/H_0^2 + 3 \Omega_{m}}{6 f' -
\displaystyle\frac{54 x \Omega_{m} H_0^2 f'' }{(x f' -
3\Omega_{\Lambda} H_0^2 f'')}}.
\end{equation}
\no Where $f(R)$ and its derivatives are as given above. Note that we can
recover the usual Friedmann equation for $f(R)=R$, the case for
which our action reduces to Einstein Hilbert action.

\vskip 0.5cm \no In this paper we test the viability of $f(R)$
cosmology. The Friedmann equation evaluated at $z=0$ imposes a
constraint on the free parameters $n$ and $\lambda$.

\vskip 0.5cm \no To check the compatibility of Starobinsky model
with observations we need to first obtain the Hubble parameter $H$,
as a function of redshift $z$. The Hubble equation in its present
form depends on the curvature $R$, through $f$ and its derivatives.
Therefore, we need to solve the trace equation (equation
(\ref{treqn})), relating curvature with redshift. In order to study
the behavior of trace equation, which does not admit a simple
solution for general $n$, we first analyse its quantitative behavior
for large  values of $z$, corresponding to $x \ll 1$. Therefore, the
trace equation can be simplified considerably and in the leading
order we recover
\begin{equation}\label{leadtrace}
    x = \frac{\Omega_{\Lambda}}{\Omega_{m0}(1+z)^3 + 2\lambda \Omega_{\Lambda}}.
\end{equation}
\no Similarly, the Friedmann equation in the leading order gives
\begin{equation}\label{leadfr}
    \frac{H^2}{H_0^2} = \Omega_{m0}(1+z)^3 + \frac{\lambda}{2}
    \Omega_{\Lambda}.
\end{equation}
\no Thus at large redshift we recover the standard $\Lambda$CDM
cosmology for $\lambda=2$, independently of $n$ (see figure \ref{fig:hzn1}).

\vskip 0.5cm \no There is no general solution for the trace equation
and therefore we analyse this model for fixed values of $n$. It was
shown by Capozziello and Tsujikawa \cite{Tsujikawa} that we need $n
> 0.9$ to satisfy solar system tests. In this work we consider model
with $n = 1$ and $2$.

\subsection{n=1}
\begin{figure}[ht]
\centering
\includegraphics[totalheight=0.5\textheight, angle=-90]{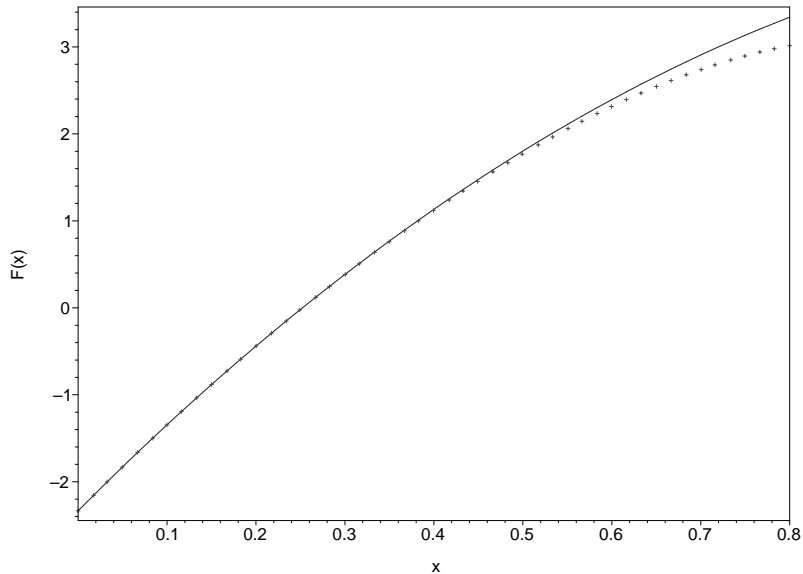}
\caption{Plot of F(x) at $z=0, \lambda =2$ (indicated by solid line)
vs truncated expression of F(x) at $z=0, \lambda=2$ (indicated by
dotted line).}\label{fig:fitn12}
\end{figure}

\no The trace equation in this case simplifies to a quintic equation
\begin{equation}\label{tracen1}
   F(x) \equiv x^5 - \frac{\Omega_{\Lambda}}{\Omega_{m}} x^4 + 2x^3 - 2
    \frac{\Omega_{\Lambda}}{\Omega_{m}} x^2 +x\left(1+ 2\lambda
    \frac{ \Omega_{\Lambda}}{\Omega_{m}} x^4 \right)-
    \frac{ \Omega_{\Lambda}}{\Omega_{m}}=0,
\end{equation}
\no where $\Omega_{m} = \Omega_{m0} (1+z)^3$. This equation cannot
be solved analytically for general $x$ in terms of other parameters.
Since $x \ll 1$, we approximate this equation with a quartic
equation. This approximation is found to be sufficiently accurate
for $x \ll 1$ (see Fig. \ref{fig:fitn12}).

\begin{figure}[ht]
\centering
\includegraphics[totalheight=0.5\textheight, angle=-90]{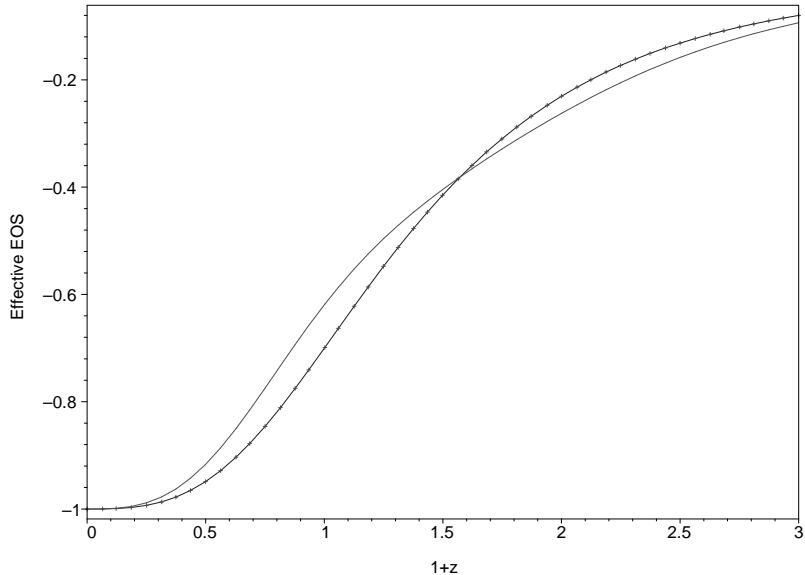}
\caption{Effective Equation of State (EOS) vs. 1+z. Solid line
corresponds to n=1 and the knotted line corresponds to
n=2. }\label{fig:eos}
\end{figure}

We need to solve for x ($x=R_0/R$) using the trace equation. We have
two quartic equations  $F(x, z, \lambda) =0$ and $G(x, z,
\lambda)=0$ for n =1 and 2 case, respectively. We shall solve the
quatric equation with $\Omega_{m0}=0.3$ and $\Omega_{\Lambda}=0.7$.
Let us first consider the case of $n =1$  and look for the roots of
$F=0$ at $z=0$. We numerically analyse the Eq. 26 and find  that for
generic values of $\lambda$  ($0<\lambda<10$), two of the four roots
of quatric equation are always imaginary.
The Friedman equation puts a constraints on the values of $\lambda$
such that $H(z)\to H_0$ for $z \to 0$, see Eq. 23. This condition
fixes the value of $\lambda$ to $2$ for one of the roots, the other
root corresponds to $\lambda \simeq 2.44$. We should now check the
viability of the roots by invoking the non-zero values f the
redshift. As z increases, i.e., we move to past, $x = R_0/R$ should
decrease  as R should increase during a viable cosmic evolution
which happens in case of $\lambda=2$. However, for the root
corresponding to $\lambda\simeq 2.44$, x increases in the past and
$x$ quickly goes beyond its normal range $0<x<1$  and we need not
consider this root at all. Therefore we have only one root obtained
at $\lambda = 2$ which produces the correct final asymptotic state
as a de Sitter model when z goes to -1 (see figure \ref{fig:eos}).
In this figure we plot the effective equation of state with
redshift:

\begin{equation}\label{eff-state}
    w_{eff} = -1 + \frac{2(1+z)}{3H}\frac{dH}{dz}.
\end{equation}
\no Once $\lambda$ is fixed there is no other free parameter in the
theory. The Hubble parameter can now be plotted for the best fit
value of $\lambda$.

\subsection{n=2}

\begin{figure}[ht]
\centering
\includegraphics[totalheight=0.5\textheight, angle=-90]{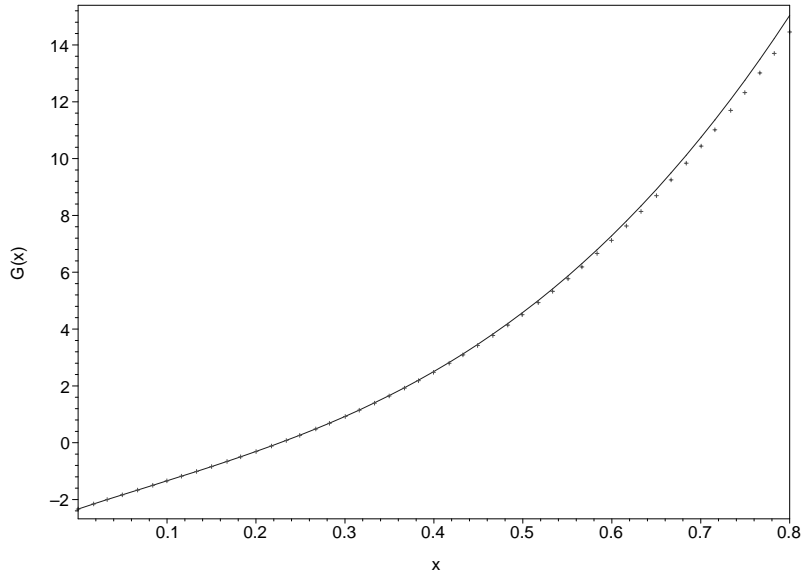}
\caption{Plot of G(x) at $z=0, \lambda =2$ (indicated by solid line)
vs truncated expression of G(x) at $z=0, \lambda=2$ (indicated by
dotted line).}\label{fig:fitn22}
\end{figure}

\begin{figure}[ht]
\centering
\includegraphics[totalheight=0.5\textheight, angle=-90]{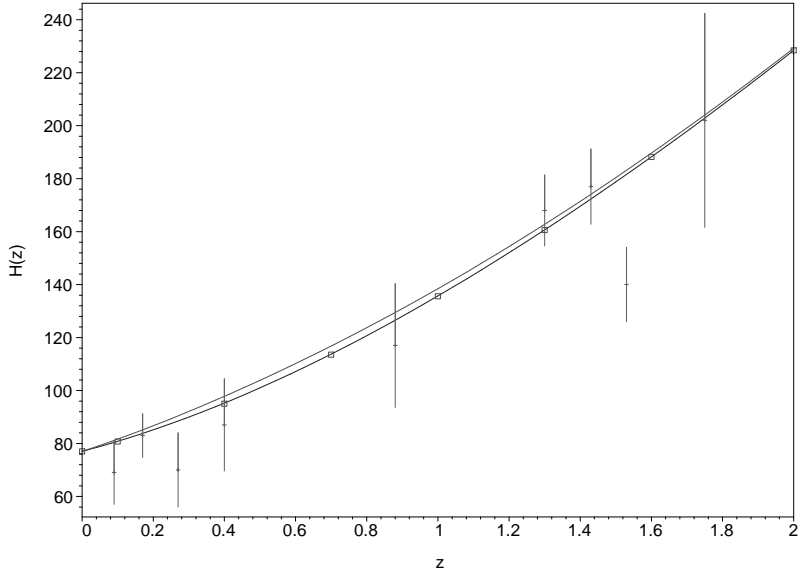}
\caption{Variation of $H(z)$ with $z$, for $H_0 = 77 \; Km s^{-1}
Mpc^{-1}$. Upper curve corresponds to n=1 case and the lower curve
is for n=2 model. The points (boxes) on the lower curve correspond
to $\Lambda$CDM.}\label{fig:hzn1}
\end{figure}

\no The trace equation in this case is a seventh degree polynomial in
$x$
\begin{equation}\label{tracen2}
 G(x) \equiv   x^7- \frac{\Omega_{\Lambda}}{\Omega_m} x^6 +3x^5 -
    3 \frac{\Omega_{\Lambda}}{\Omega_m} x^4 + 3\left(1+
    2\lambda \frac{\Omega_{\Lambda}}{\Omega_m} \right) x^3
    -3\frac{\Omega_{\Lambda}}{\Omega_m} x^2 + \left(1+
    2\lambda \frac{\Omega_{\Lambda}}{\Omega_m} \right) x -
    \frac{\Omega_{\Lambda}}{\Omega_m}=0.
\end{equation}
\no Again we can approximate this equation by a quartic equation in
$x$ using the fact that $x \ll 1$ (see Fig.\ref{fig:fitn22}). In
order to calculate the roots of this equation, we followed the same
procedure as described for $ n = 1$ case. Here also $\lambda=2$
reproduces the correct late time behavior (see figure
\ref{fig:eos}).
\section{Observational Constraints}

\no In this paper, we check the compatibility of the model with the
$H(z)$ data and the measurements of baryon acoustic oscillation
peak. We find the values of $\chi^2$ to see how well the $f(R)$
model with $n=1$ and $n =2$ accommodate the observations. Here, for
a given model, we calculate $\chi^2$ using H(z) data, BAO
measurement and for the joint data (H(z) + BAO).

\subsection*{Constraints from H(z) data}

\no Simon, Verde and Jimenez (2005) employed differential ages of
passively evolving galaxies to get the Hubble parameter as a
function of redshift, $H(z)$ \cite{svj} . The nine data points of
H(z) with $0.09 \le z \le 1.75$ have been obtained by using absolute
ages of 32 galaxies taken from the Gemini Deep Deep Survey (GDDS)
and the archival data.

\vskip 0.5cm
\no For the adopted $f(R)$ model, we calculate the values of $\chi^2$ for
$ n =1$ and $n = 2$. For this we define

\begin{equation}
\chi^2 (H_0,n)= \,\sum_{i=1}^9 { \frac{\left (
H_{\mathrm{exp}}(z_i,n,H_0) - H_{\mathrm{obs}}(z_i)
\right)^2}{\sigma_i^2}}. \label{eq:Hz}
\end{equation}

\vskip 0.5cm \no Here $H_0$ is the present day value of the  Hubble
constant, $H_{\mathrm{exp}}(z_i, n, H_0)$ is the expected value of
the Hubble constant in the $f(R)$ cosmology at redshift $z_i$ for a
particular $n$ and $H_0$. $H_{\mathrm{obs}}$ is the observed value
and $\sigma_i$ is the corresponding $1\,\sigma$ uncertainty in the
measurement. The sum is over all observed data points (nine in
number).

\vskip 0.5cm
\no As the value of $\chi^2$ is highly sensitive to the value of $H_0$,
we marginalize over $H_0$ to obtain the modified $\chi^2$. For this
we define the likelihood function as:

\[
\mathcal{L} = \int \, e^{-\chi^2/2}\,P(H_0)\,\, dH_0.
\]
\no Here $P(H_0)$ is the prior probability function for $H_0$ which is
Gaussian:

\[
P(H_0) = \frac{1}{\sqrt{2\pi} \sigma_{H_0}} \;
\exp\left[-\frac{1}{2}
  \; \frac{(H_0-
H_0^{\mathrm{obs}})^2}{\sigma_{H_0}^{2}}\right],
\]

\no with $H_0^{\mathrm{obs}}$ as the value of $H_0$ (and $\sigma_{H_0}$
is the error in it) as suggested by independent observations. In
this paper, we also study the effect of different priors on the
result. We use two set of priors:

\vskip 0.3 cm

\no {\bf Set A:} $ H_{0}^{\mathrm{obs}}\,\,=\, 68 \pm 4$ Km/s/Mpc,
as obtained from the median statistics analysis of 461 measurements
of $H_0$ \cite{chen}. \vskip 0.3 cm \no {\bf Set B:}
$H_{0}^{\mathrm{obs}}\,\, =77 \pm 4$ Km/s/Mpc, as suggested by the
Chandra  X - ray  Observatory results \cite{cxo}. \vskip 0.4cm
\vskip 0.3cm \no We observe that the value of the modified
$\chi^2_{H(z)}$ is sensitive to the choice of the prior.

\subsection* {Baryon Acoustic Oscillation (BAO)}
\no Before recombination, the universe was in the completely ionized
state. The cosmological perturbations in the relativistic primordial
baryon-photon plasma produced acoustic oscillations. These
oscillations were imprinted in the form of peaks in the late time
power spectrum of non-relativistic matter. This acoustic peak which
is predicted at the measured scale of $ 100 \, h^{-1} \, Mpc$ is
detected in the large scale correlation function of 46478 sample of
luminous red galaxies in the Sloan Digital Sky Survey\cite{Eis}. It is
described by a dimensionless parameter $\cal {A}$,

\begin{equation}
{\cal{A}}(n) \equiv
\frac{\Omega_{\rm{m}}^{1/2}}{z_{\rm{*}}}\left[z_{\rm{*}}
  \frac{\Gamma^{2}(z_{\rm{*}},n)}{{\cal{E}}(z_{\rm{*}},n)}\right]^{1/3}
\end{equation}
\noindent where $z_{\rm{*}} = 0.35$, $\Gamma(z_{\rm{*}},n) =
\int_0^{z_{\rm{*}}}dz/{\cal{E}}(z_{\rm{*}},n)$ is the dimensionless comoving
distance to $z_{\rm{*}}$, ${\cal{E}}(z_{\rm{*}},n)$ is
given by $H(z)/H_0$.

\vskip 0.5cm
\no We define $\chi^2_{\mathrm BAO}= {{({\cal {A}}(n) -
A_{\mathrm{obs}})}^2}/{\sigma_{\cal{A}}^2}$, with $\mathcal{A}_{\mathrm{obs}} = 0.469 \pm 0.017$. The values of $\chi^2_{\mathrm BAO}$ for $ n = 1$ and $n = 2$ are listed in Table 1.

\begin{table}[ht]

\begin{center}
\begin{tabular}{l l l l l r}\hline\hline
n & $H_0$ prior & {\bf $ \chi^{2}_H(z)$}& {\bf
$\chi^2_{\mathrm{BAO}}$}&
{\bf$\chi^2_{\mathrm{Total}}$}& {\bf$\chi^{2}_{\nu}$}\\
\hline \hline
&&&&& \\
1& $68 \pm 4$& 7.29& 0.54 & 7.83 &0.87\\
1& $77\pm 4$& 6.63&0.54 & 7.17& 0.8\\
&&&&&\\
2& $68\pm 4$&7.79& 0.88 & 8.67 & 0.96\\
2& $77\pm 4$&6.30 & 0.88 & 7.18 & 0.8\\
&&&& &\\
\hline
\end{tabular}
\caption{Here $\chi^2_{H(z)} = -\frac{1}{2}\,\,\ln(\mathcal{L})$.}
\end{center}

\end{table}

\section{Results and Discussion}

\no The observed late time acceleration of the universe is one of the
major unsolved problem in the cosmology. It may hint at the breakdown
of Einstein GR . This has lead to modification of the Einstein theory of
gravity. One of the attractive possibility to modify this theory is to
replace the Ricci scalar $R$ with the generic function $f(R)$ in the
Hilbert action.

\vskip 0.5 cm
\no $f(R)$ theories are usually studied by two methods: {\it metric} and
{\it Palatini} approach. The Palatini approach is explored
extensively in the literature both theoretically and
observationally \cite{pal}( see the other ref's given in these
papers). This formulation gives second order differential
field equation which can explain the late time behavior of the
universe.

\vskip 0.5 cm \no The metric approach of $f(R)$ theory  leads to a
fourth order non linear differential equation in terms of scale
factor. This equation is difficult to solve both analytically and
numerically even for the special cases. Therefore, not much
observational tests have been performed on $f(R)$ theories based on
metric formulation. Our work is an attempt to check the concordance
of $f(R)$ theory of gravity using the metric approach  with some of
the cosmological observations. In particular, we have explored  the
Starobinsky model in which function $f(R)$ is analytic, satisfying
the condition $ f(0) = 0$. This model also passes the Solar system
and laboratory tests successfully for large values of $n$. In this
work we study the $f(R)$ model for $n= 1$ and $n=2$. \vskip 0.5 cm
\noindent In order to study the cosmological viability of this
theory, we investigate this class of model with two observational
tests. The first method is based on the Hubble parameter versus
redshift data, $H(z)$. The Hubble parameter  is related to the
differential age of the universe through this form
$$
H(z) = - \frac{1}{1 +z} \frac{dz}{dt}.
$$
 By estimating the $dt/dz$, one can obtain directly the Hubble
parameter, $H(z)$ at different redshifts. The H(z) data has one
major advantage, unlike in the standard candle approach (SNe Ia):
the Hubble function is not integrated over. The other important
feature of this test is that differential ages are less sensitive to
systematic errors as compared to the absolute ages \cite{svj}. This
observational $H(z)$ have been used earlier also to constrain
various other dark energy models \cite{other}.

\vskip 0.5 cm \noindent In this work, we check the compatibility of
$f(R)$ model with $n =1$ and $n=2$ with the $H(z)$ data. Since $H_0$
is a nuisance parameter, we marginalise over $H_0$. We further
combine the results obtained from $H(z)$ data set with the BAO data.
To perform the joint test, we define the quantity:
\begin{equation}
\chi^2_{\mathrm{total}} = \chi^2_{\mathrm{H(z)}} + \chi^2_{\mathrm{BAO}}.
\end{equation}

\no The results are given in the Table 1. It appears that $n = 2$
is favored by the observations. This fact is also in agreement with
solar and laboratory test. The other important conclusions are as
follow:

\begin{itemize}
\item \no The variation of $H(z)$ with redshift becomes independent of
$n$ after the redshift around $ z =1.8$, see Fig. 7. Before this
redshift ( $ z < 1.8$),  there is a small difference between the
behavior of $H(z)$ with $z$ for $ n =1$ and $n =2$ models. This
variation is still within the error bars of the data points. At
higher redshift ( i.e when $R >> R_0$) we recover the standard
$\Lambda$CDM universe for $\lambda =2$, for all the values of $n$.
Hence the thermal history of the universe is correctly reproduced by
this model.

\item \no The expansion history for this model with $ n = 2,
\lambda = 2$ matches exactly with the standard cosmological model
($\Lambda$CDM) with equation of state parameter $\omega = -1,
\Omega_m = 0.3$ and $\Omega_{\Lambda} = 0.7$ (see fig. 7). The
variation in parameters $\Omega_m$ and $\Omega_{\Lambda}$ (keeping
$\Omega_{m0} + \Omega_{\Lambda} = 1$) leads to very small changes in
values of $\lambda$, keeping it close to 2.

\item \no As shown in Table 1, it seems
that the present observational data used in this work prefer $n = 2$
over $n = 1$, as indicated by $\chi^{2}$ per degree of freedom,
$\chi^2_{\nu}$. However, $ n=1$ cannot be ruled out.

\item \no As stated
earlier, the algebraic expressions become cumbersome for larger
values of $n$. It is quite possible that observational tests
discussed in this paper may be compatible with values of $n$ larger
than two which are also consistent with solar and laboratory tests.
However, it should be emphasized that for a given value of
$\lambda$, the minimum of the scalaron potential gets closer to zero
for larger values of $n$ making the model vulnerable to singularity,
see Fig. 3.

\end{itemize}

\vskip 0.5cm \no Due to large error bars in the data sets used, the
variation of the expansion history of the universe studied in this
paper can easily be accommodated by the observations, thereby
making the models close to $\Lambda$CDM at the background level.
\no At present the sample of $H(z)$ data is too small  and the error
bars are large. In the future, large amount  of precise $H(z)$
data is expected to become available. This will not only  reveal the
fine features of the expansion history of the universe but also tightly
constrain the cosmological parameters.

\vskip 0.5cm \no There is a further need to explore this model with
time based observational tests, like age of the universe and high
redshift objects. It is  known that the evolution of the age of the
universe with the redshift vary from model to model. So it is
possible that the model of the universe which are able to reproduce
the total age of the universe at $z= 0$, may not accommodate the
objects at high redshift\cite{dj}. Therefore the time based
observational test may  play the key role in short-listing the
viable dark energy model in the near future. It is also expected
that the study of matter power perturbations would allow to
distinguish this model from the standard cosmology.

\section*{Acknowledgments}
We thank V. Sahni and S. Tsujikawa for useful comments. AD $\&$ DJ  thank
A. Mukherjee and S. Mahajan for providing  the facilities
to carry out the research work. This work is supported by
Japanese-Indian collaboration project (Grant No
DST/INT/JSPS/Project-35/2007). The research by SN has been supported
in part by the Monbu-Kagaku-sho of Japan under grant no.18549001 and
Global COE Program of Nagoya University provided by the Japan Society
for the Promotion of Science (G07).

\end{document}